\documentclass[prl,showpacs,amsmath,amssymb,superscriptaddress]{revtex4}

\usepackage[dvips]{graphicx}

\usepackage{graphics}

\usepackage{dcolumn}

\usepackage{bm}

\usepackage[usenames]{color}

\begin{document}

\title[Lithographically and electrically controlled strain effects on AMR in (Ga,Mn)As]{Lithographically and electrically controlled strain effects on anisotropic magnetoresistance in (Ga,Mn)As}
\author{E. De Ranieri}
\affiliation{Microelectronics Research Centre, Cavendish Laboratory,
University of Cambridge, CB3 0HE, UK}
\affiliation{Hitachi Cambridge Laboratory, Cambridge CB3 0HE, United Kingdom}

\author{A.~W.~Rushforth}
\affiliation{School of Physics and Astronomy, University of Nottingham, Nottingham NG7 2RD, United Kingdom}

\author{K. V\'yborn\'y}
\affiliation{Institute of Physics ASCR v.v.i., Cukrovarnick\'a 10, 162 53 Praha 6, Czech Republic}

\author{U.~Rana}
\affiliation{Microelectronics Research Centre, Cavendish Laboratory,
University of Cambridge, CB3 0HE, UK}
\affiliation{Hitachi Cambridge Laboratory, Cambridge CB3 0HE, United Kingdom}

\author{E.~Ahmed}
\affiliation{School of Physics and Astronomy, University of Nottingham, Nottingham NG7 2RD, United Kingdom}

\author{R.~P.~Campion}
\affiliation{School of Physics and Astronomy, University of Nottingham, Nottingham NG7 2RD, United Kingdom}

\author{C.~T.~Foxon}
\affiliation{School of Physics and Astronomy, University of Nottingham, Nottingham NG7 2RD, United Kingdom}

\author{B.~L.~Gallagher}
\affiliation{School of Physics and Astronomy, University of Nottingham, Nottingham NG7 2RD, United Kingdom}

\author{A.~C.~Irvine}
\affiliation{Microelectronics Research Centre, Cavendish Laboratory,
University of Cambridge, CB3 0HE, UK}

\author{J.~Wunderlich}
\affiliation{Hitachi Cambridge Laboratory, Cambridge CB3 0HE, United Kingdom}
\affiliation{Institute of Physics ASCR v.v.i., Cukrovarnick\'a 10, 162 53 Praha 6, Czech Republic}

\author{T.~Jungwirth}
\affiliation{Institute of Physics ASCR v.v.i., Cukrovarnick\'a 10, 162 53 Praha 6, Czech Republic}
\affiliation{School of Physics and Astronomy, University of Nottingham, Nottingham NG7 2RD, United Kingdom}

\date{\today}

\begin{abstract}
It has been demonstrated that magnetocrystalline anisotropies in (Ga,Mn)As are sensitive to lattice strains as small as $10^{-4}$ and that strain can be controlled by lattice parameter engineering during growth, through post growth lithography, and electrically by bonding the (Ga,Mn)As sample to a piezoelectric transducer. In this work we show that analogous effects are observed in crystalline components of the anisotropic magnetoresistance (AMR). Lithographically or electrically induced strain variations can produce crystalline AMR components which are larger than the crystalline AMR and a significant fraction of the total AMR of the unprocessed (Ga,Mn)As material. In these experiments we also observe new higher order terms in the phenomenological AMR expressions and find that strain variation effects can play important role in the micromagnetic and magnetotransport characteristics of (Ga,Mn)As lateral nanoconstrictions.

\end{abstract}

\pacs{75.47.-m, 75.50.Pp, 75.70.Ak}

\maketitle


\section{1. Introduction}
GaAs doped with $\sim 1-10$\% of the magnetic acceptor Mn is a unique material for exploring spin-orbit coupling effects on micromagnetic and magnetotransport characteristics of ferromagnetic spintronic devices.  Spin polarized valence band holes that mediate ferromagnetic coupling between  Mn local moments produce large magnetic stiffness, resulting in a mean-field like magnetization and macroscopic single-domain behavior of these dilute moment ferromagnets. At the same time, magnetocrystalline anisotropies derived from spin-orbit coupling effects in the hole valence bands are large leading to the sensitivity of magnetic state to strains as small as $10^{-4}$ \cite{Humpfner:2006_a,Wunderlich:2007_c,Wenisch:2007_a,Rushforth:2008_a,Overby:2008_a}. Experimentally, strain effects can be controlled by lattice parameter engineering  during growth \cite{Dietl:2001_b,Abolfath:2001_a}, through post growth lithography \cite{Humpfner:2006_a,Wunderlich:2007_c,Wenisch:2007_a}, or electrically by bonding the (Ga,Mn)As sample to a piezoelectric transducer \cite{Rushforth:2008_a,Overby:2008_a,Goennenwein:2008_a}. Easy axis rotations from in-plane to out-of-plane directions have been demonstrated in these studies in  (Ga,Mn)As films grown under compressive and tensile lattice matching strains, and the orientation of the in-plane easy axis (axes) has been shown to respond to strain relaxation in lateral microstructures or controlled dynamically by piezoelectric transducers.

Strain control of magnetocrystalline effects on transport in (Ga,Mn)As, we focus on in this paper, has so far been explored less extensively. Our work in this direction is motivated by previous experimental and theoretical analyses of the  crystalline terms of the anisotropic magnetoresistance (AMR). The studies have shown that these magnetotransport coefficients can be large and reflect the rich magnetocrystalline anisotropies of the studied (Ga,Mn)As materials \cite{Baxter:2002_a,Jungwirth:2003_b,Tang:2003_a,Matsukura:2004_a,Goennenwein:2004_a,Wang:2005_c,Limmer:2006_a,Rushforth:2007_a,Rushforth:2007_b}. Furthermore, the possibility of directly controlling the crystalline AMR by strain has been demonstrated in (Ga,Mn)As films grown under compressive and tensile lattice matching strains \cite{Jungwirth:2002_c,Jungwirth:2003_b,Matsukura:2004_a}. Here we report and analyze AMR measurements in strain relaxed (Ga,Mn)As micro and nanostructures and in a (Ga,Mn)As film bonded to a piezo-stressor. We show that post-growth induced lattice distortions can significantly modify crystalline AMR terms  and give rise to new, previously undetected components in phenomenological AMR expansions.

The paper is organized as follows: In the 2nd section we derive a phenomenological description of the AMR relevant to the experimentally studied (Ga,Mn)As systems. 3rd and 4th sections report AMR data acquired in macroscopic and strain-relaxed microscopic Hall bars, in (Ga,Mn)As nanoconstriction devices, and in  voltage controlled piezoelectric/(Ga,Mn)As hybrid structures. A brief summary of the results is given in the 5th Section.

\section{2. Phenomenological description of the AMR}

We consider a thin film geometry and a magnetization vector, $\vec{M}/|\vec{M}|=(\cos\psi,\sin\psi)$, in the plane of the film with its two components defined with respect to the orthogonal crystallographic basis $\{[100],[010]\}$.
The resistivity tensor,
\begin{equation}
  \hat{\rho} = \left(\begin{array}{cc}
     \rho_{11}(\cos\psi,\sin\psi) & \rho_{12}(\cos\psi,\sin\psi) \\
     \rho_{21}(\cos\psi,\sin\psi) & \rho_{22}(\cos\psi,\sin\psi)
               \end{array}\right)\; ,
\label{eq-01}
\end{equation}
written in the same basis describes the longitudinal and transverse resistivities of a pair of Hall bar
devices oriented along the $[100]$-direction ($\rho_{11}$ and $\rho_{21}$) and the $[010]$-direction ($\rho_{22}$ and $\rho_{12}$). Resistivities of a pair of orthogonal Hall bars tilted by an angle $\theta$ from  the [100]/[010] directions are given by $R_{-\theta} \hat{\rho}R_{\theta}$, where $R_\theta=\left(\begin{array}{cc}
     \cos\theta & -\sin\theta \\
     \sin\theta & \cos\theta
               \end{array}\right)$ is the rotation matrix. Written explicitly, the longitudinal ($\rho_L$) and transverse ($\rho_T$) resistivities for the Hall bar rotated by the angle $\theta$ from the [100]-direction read
\begin{equation}
\begin{array}{rcl}
\rho_L &=&
(\cos\theta,\sin\theta)\cdot \hat{\rho}\cdot \displaystyle {\cos\theta \choose \sin\theta}\,, \\ \\
\rho_T &=&
(\cos\theta,\sin\theta)\cdot \hat{\rho}\cdot \displaystyle {-\sin\theta \choose \cos\theta}\,.
\end{array}
\label{eq-02}
\end{equation}

We first derive expressions for the non-crystalline AMR components \cite{McGuire:1975_a,Rushforth:2007_a} which depend only on the angle $\psi-\theta$ between the current (Hall bar orientation) and the magnetization vector, and which account for the AMR in isotropic (polycrystalline) materials. We expand the elements of $\hat{\rho}$ in Eq.~(\ref{eq-01}) in series of $\cos^n\psi$ and $\sin^n\psi$, or equivalently of $\cos n\psi$ and $\sin n\psi$ \cite{McGuire:1975_a}. The form of Eq.~(\ref{eq-02}) implies that corresponding expansions of $\rho_L$ and $\rho_T$   in series of $\cos(n\psi+m\theta)$ and $\sin(n\psi+m\theta)$ contain only terms with $m=0,\pm 2$. Among those, the $\cos2(\psi-\theta)$ and $\sin2(\psi-\theta)$ are the only terms depending on
$\psi-\theta$. It explains why the non-crystalline AMR components, which are obtained by truncating  Eq.~(\ref{eq-01}) to
\begin{equation}
\hat{\rho} = \rho_{av}\left(\begin{array}{cc}1 &0\\ 0&1 \end{array}\right) +
         2\rho_{av}C_I \left(\begin{array}{cc}
             -\frac{1}{2}+\cos^2\psi & \sin\psi\cos\psi \\
             \sin\psi\cos\psi & -\frac{1}{2}+\sin^2\psi \end{array}\right)\,,
\label{eq-03}
\end{equation}
take the simple form $\Delta\rho_L/\rho_{av}\equiv (\rho_L-\rho_{av})/\rho_{av} = C_I\cos 2(\psi-\theta)$ and $\rho_T/\rho_{av} = C_I\sin 2(\psi-\theta)$ \cite{McGuire:1975_a,Rushforth:2007_a}. Here $\rho_{av}$ is the average (with respect to $\psi$) longitudinal resistivity, and $C_I$ is the non-crystalline AMR amplitude.

All terms in the expansion of Eq.~(\ref{eq-02}) which depend explicitly on the orientation of the magnetization vector with respect to the crystallographic axes contribute to the crystalline AMR \cite{McGuire:1975_a,Rushforth:2007_a}. Symmetry considerations can be used to find the form of $\hat{\rho}(\psi)$ in Eq.~(\ref{eq-01}) specific to a particular crystal structure. Explicit expressions for $\hat{\rho}(\psi)$ in unperturbed cubic crystals, and cubic crystals with uniaxial strains along [110] and [100] axes are derived in the Appendix. Here we write the final expression for $\rho_L$ and $\rho_T$ obtained from the particular form of $\hat{\rho}(\psi)$ and from Eq.~(\ref{eq-02}). For the cubic lattice, omitting terms with the periodicity in $\psi$ smaller than $90^\circ$, we obtain,
\begin{eqnarray}\label{eq-04a}
  \displaystyle\frac{\Delta\rho_L}{\rho_{av}} &=&
C_I \cos 2(\psi-\theta) + C_{IC}\cos (2\psi+2\theta) + C_C \cos 4\psi+\ldots \\
\label{eq-04b}
  \displaystyle\frac{\rho_T}{\rho_{av}} &=&
C_I \sin 2(\psi-\theta) - C_{IC}\sin (2\psi+2\theta) + \ldots \;.
\end{eqnarray}
For the higher order cubic terms see the Appendix.

Additional components emerge in $\Delta\rho_L/\rho_{av}$ and $\rho_T/\rho_{av}$ for the uniaxially strained lattice which we denote as $\Delta_L^{uni}$ and  $\Delta_T^{uni}$, respectively. Omitting terms with the periodicity in $\psi$ smaller than $180^\circ$ we obtain,
\begin{equation}\label{eq-11a}\begin{array}{rcl}
  (\pm)\Delta_L^{uni} &=& C_{IU}^{s}\sin 2\theta + C_U^{s}\sin 2\psi \\
  (\pm)\Delta_T^{uni} &=& C_{IU}^{s}\cos 2\theta
\end{array}
\end{equation}
for strain along the in-plane diagonal directions ($s=[110]$ corresponds to "+" and $[1\bar{1}0]$ to "-"), and
\begin{equation}\label{eq-11b}\begin{array}{rcl}
  (\pm)\Delta_L^{uni} &=& C_{IU}^{s}\cos 2\theta + C_U^{s}\cos 2\psi \\
  (\pm)\Delta_T^{uni} &=& - C_{IU}^{s}\sin 2\theta + C_{U,T}^{s}\sin 2\psi
\end{array}
\end{equation}
for strain along the in-plane cube edges ($s=[100]$ corresponds to "+" and $[010]$ to "-") For higher order uniaxial terms see again the Appendix.

\section{3. Experiments in lithographically patterned (Ga,Mn)As microdevices}
\begin{figure}[ht]
\vspace{1cm}
\hspace{0cm}\includegraphics[width=.8\columnwidth,angle=90]{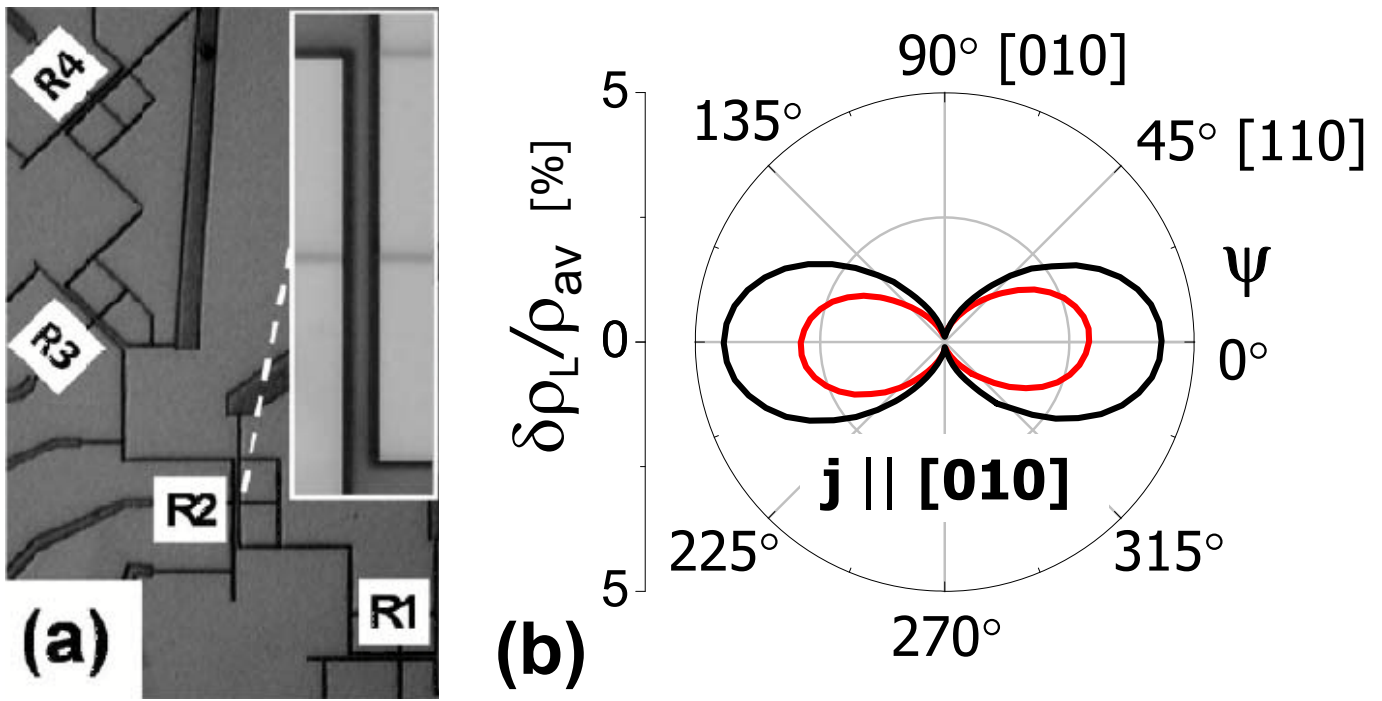}
\vspace*{-2cm}
\caption{(a) Micrograph of the first studied device with four
microscopic bars in series (the macroscopic Hall bar is not shown). Inset:
enlarged view of one of the microchannels; black areas are the isolating
trenches defining the channel. (b) Polar plot of the percentage change
in resistivity (AMR)
as a function of the angle between the applied field and the
[100] direction for the micro- and macroscopic bars aligned along the [010]
axis. Strain relaxation due to patterning leads to a reduction
of the AMR magnitude of about 30\%. For better clarity we plot, instead of $\Delta\rho_L/\rho_{av}$ defined in the Section~2, $\delta\rho_L/\rho_{av}\equiv (\rho_L-\rho_{L,min})/\rho_{av}$. Here $\rho_{L,min}$ is the minimum (with respect to $\psi$) longitudinal resistivity.}
\label{f1}
\end{figure}

We now proceed with the discussion of AMR measurements in (Ga,Mn)As microdevices in which strain effects are controlled by lithographically induced lattice relaxation \cite{Humpfner:2006_a,Wunderlich:2007_c,Wenisch:2007_a}.
Optical micrograph of the first studied device is shown in Fig.~\ref{f1}(a).
The structure consists of four 1~$\mu$m wide Hall bars and one 40~$\mu$m wide
bar connected in series. The wider bar is aligned along the [010]
crystallographic direction, the micro-bars are oriented along the [110],
[1$\overline{1}$0], [100], and [010] axes. The Hall bars are defined by 500~nm
wide trenches patterned by e-beam lithography and reactive ion etching in a
25~nm thick Ga$_{0.95}$Mn$_{0.05}$As epilayer, which was grown along the [001]
crystal axis on a GaAs substrate. The Curie temperature of the as-grown
(Ga,Mn)As is 60~K. A compressive strain in the (Ga,Mn)As epilayer grown on the
GaAs substrate leads to a strong magnetocrystalline anisotropy which forces
the magnetization vector to align parallel with the plane of the magnetic
epilayer \cite{Dietl:2001_b,Abolfath:2001_a}. The growth strain is partly relaxed in the microbars, producing an additional, in-plane uniaxial tensile strain in the transverse direction \cite{Humpfner:2006_a,Wunderlich:2007_c,Wenisch:2007_a}.

Magnetoresistance traces were measured with the saturation magnetic field
applied in the plane of the device, {\em i.e.}, in the pure AMR geometry with
zero (antisymmetric) Hall signal and with magnetization vector aligned with
the external magnetic field. The sample was rotated by 360$^{\circ}$ with
5$^{\circ}$ steps. Longitudinal resistances of all five Hall-bars were
measured simultaneously with lock-in amplifiers.

In Fig.~\ref{f1}(b) we show AMR data from magnetization rotation experiments
in the 40~$\mu$m and 1~$\mu$m wide bars aligned along the [010] direction. Both
curves have a minimum for magnetization oriented parallel to the Hall bar axis
and a maximum when magnetization is rotated by 90$^{\circ}$. Although this is
a typical characteristic of the non-crystalline AMR term in (Ga,Mn)As the
large difference between the AMR magnitudes in the two devices points to a
strong  contribution of the crystalline AMR coefficient $C_U^{[010]}$ in Eq.~(\ref{eq-11b}), originating from the strain induced by transverse lattice relaxation in the microbar. We find that the magnitude of the coefficient, $C_U^{[010]}=0.77$, amounts to about 30\% of the magnitude of the total AMR in the
unrelaxed macroscopic bar.

\begin{figure}[h]
\hspace{0cm}\includegraphics[width=.8\columnwidth,angle=90]{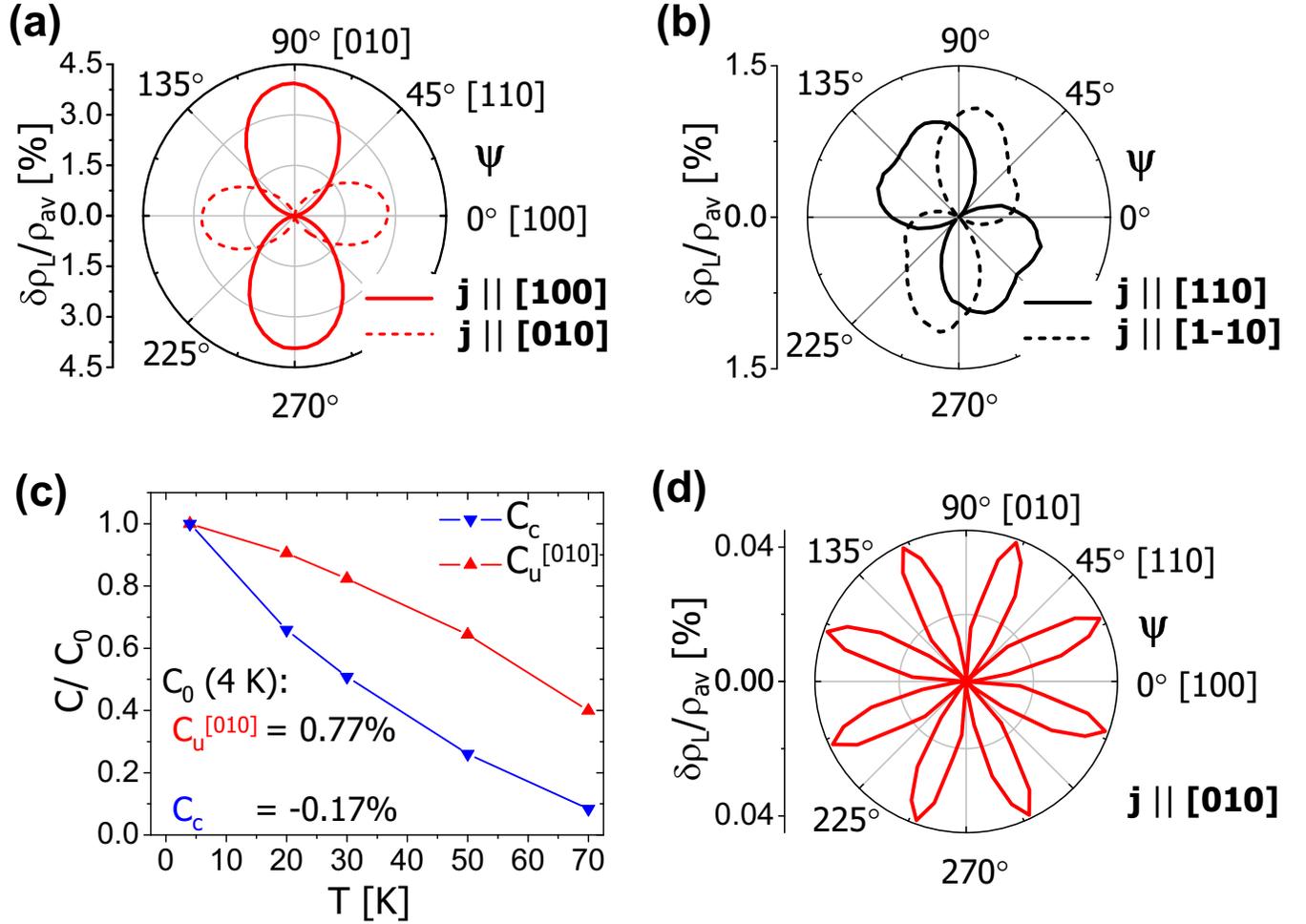}

\caption{(a),(b) AMR curves for the microscopic bars aligned along the in-plane
cubic and diagonal directions respectively. The magnitude of the AMR
is about a factor of 3 smaller in the microbars along the diagonal directions.
(c) Temperature dependence of the crystalline AMR coefficients
normalized to the respective values $C_0$ at 4 K (shown in the inset). (d)
Polar plot showing the 8th order term found in the AMR of the microscopic
bars, with a magnitude of 0.04\%.}
\label{f2}
\end{figure}

In Figs.~\ref{f2}(a) and (b) we plot AMR traces for microbars patterned
along the [100]/[010] and [110]/[1$\bar{1}$0] crystallographic directions. Strikingly, the overall magnitude of the AMR traces for the [110]/[1$\bar{1}$0] oriented Hall bars is about a factor of 3 smaller than for the [100]/[010] bars and appears to have a much stronger relative contribution of the cubic crystalline term (the term proportional to $C_c$ in Eq.~(\ref{eq-04a})). However, by extracting the 90$^{\circ}$-periodic AMR components for all microbars, as well as for the macroscopic Hall bar, we find a consistent value of $C_c=-0.17\pm 0.01$\%. This implies that it is rather a suppression (enhancement) of the  uniaxial AMR components for the [110]/[1$\bar{1}$0] ([100]/[010]) oriented bars  which accounts for the difference in AMR traces in Figs.~\ref{f2}(a) and (b).
Since $\theta=n\times 45^{\circ}$ for the Hall bars studied in Figs.~\ref{f1} and \ref{f2} and the lattice relaxation induced strains in these microbars are in the transverse direction we can rewrite the $\psi$-dependent uniaxial terms for the longitudinal AMR in Eqs.~(\ref{eq-11a},\ref{eq-11b}) in a compact form,
\begin{equation}
  \Delta_{L}^{uni} = C_{U}^{s} \cos2(\psi-\theta)\;.
\label{eq-05}
\end{equation}
This expression, together with Eq.~(\ref{eq-04a}), implies that
the amplitude of the total uniaxial (180$^{\circ}$-periodic) contribution to the AMR in the [110]/[1$\bar{1}$0] devices
($|C_I+C_{U}^{[100]/[010]}+C_{IC}|$) can indeed differ from that of the [100]/[010] devices
($|C_I+C_{U}^{[110]/[1\bar{1}0]}-C_{IC}|$), provided that $C_{IC}$ is non-zero and/or $C_{U}^{[100]/[010]}\neq C_{U}^{[110]/[1\bar{1}0]}$.

Another observation we make  is a broken [100]-[010] symmetry between the two AMR traces in Fig.~\ref{f2}(a) and in each of the two traces in Fig.~\ref{f2}(b). While in the former case this behavior can be captured by Eq.~(\ref{eq-05}) taking $C_{U}^{[100]}\neq C_{U}^{[010]}$, the shape of the AMR curves in Fig.~\ref{f2}(b) is inconsistent with the form of Eq.~(\ref{eq-05}). We have attempted to model the broken [100]-[010] symmetry by introducing a contribution to the $C_{U}^{[100]}$ coefficient which is independent of the microbar orientation, {\em i.e.}, assuming that its origin is distinct from transverse strains induced by the micropatterning. From the difference between the two AMR curves in Fig.~\ref{f2}(a) and from the [1$\bar{1}$0]-bar AMR in Fig.~\ref{f2}(b) we obtained that this contribution is 0.3\%, and from the [110]-bar AMR we obtained 0.1\%. A bar independent contribution to $C_{U}^{[100]}$ therefore explains only part of the observed [100]-[010] broken symmetry effects; we attribute the remaining part to possible material inhomogeneities or non-uniformities and misalignments in the micropatterning.

Importantly, the above experimental uncertainties have no effect on the main conclusion of our experiments that the lattice relaxation induced uniaxial AMR coefficient is larger than the cubic crystalline component and a significant fraction of the total AMR of the unpatterned material. By normalizing the value of the transverse strain-induced $C_{U}^{[010]}$ coefficient and the $C_{C}$ coefficient to the
respective values at 4~K, we can also compare their temperature dependencies within
the measured range of temperatures of 4 to 70~K. Clearly the $C_{C}$ coefficient decreases more rapidly
with increasing temperature. This recalls the behavior of the
magnetocrystalline anisotropy terms in magnetization, where the uniaxial term decreases in a
less pronounced way than the cubic one, since the former scales roughly with $M^2$, while the latter with $M^4$. As a result of
this, the transverse strain-induced term becomes more dominant at higher temperatures,
changing from 31\% of the total AMR at 4 K, to 38\% at 70 K (not shown). (Note
that non-zero AMR is still observable at 70~K which is above the $T_{c}$ of
the as-grown material. This is presumably because of the increase in the Curie
temperature due to partial annealing during the device fabrication processes.)

A detail analysis of the longitudinal resistance measurements in the microbars  allows us
to identify higher order cubic terms (see Eq.~(\ref{eq-08a}) in the Appendix).  By subtracting the 2nd and 4th order terms from AMR data measured on the [010] microbar we find a clear signature of an 8th order ($45^\circ$-periodic) cubic term with an amplitude of 0.04\%, as shown in Figure~\ref{f2}(d). In Sections~4 we give another example of the unusual high order AMR terms (and explain in more detail how these are extracted from the data) which emerge from post-growth induced lattice distortion experiments.

\begin{figure}[h]
\vspace{1cm}
\hspace{0cm}\includegraphics[width=.8\columnwidth,angle=90]{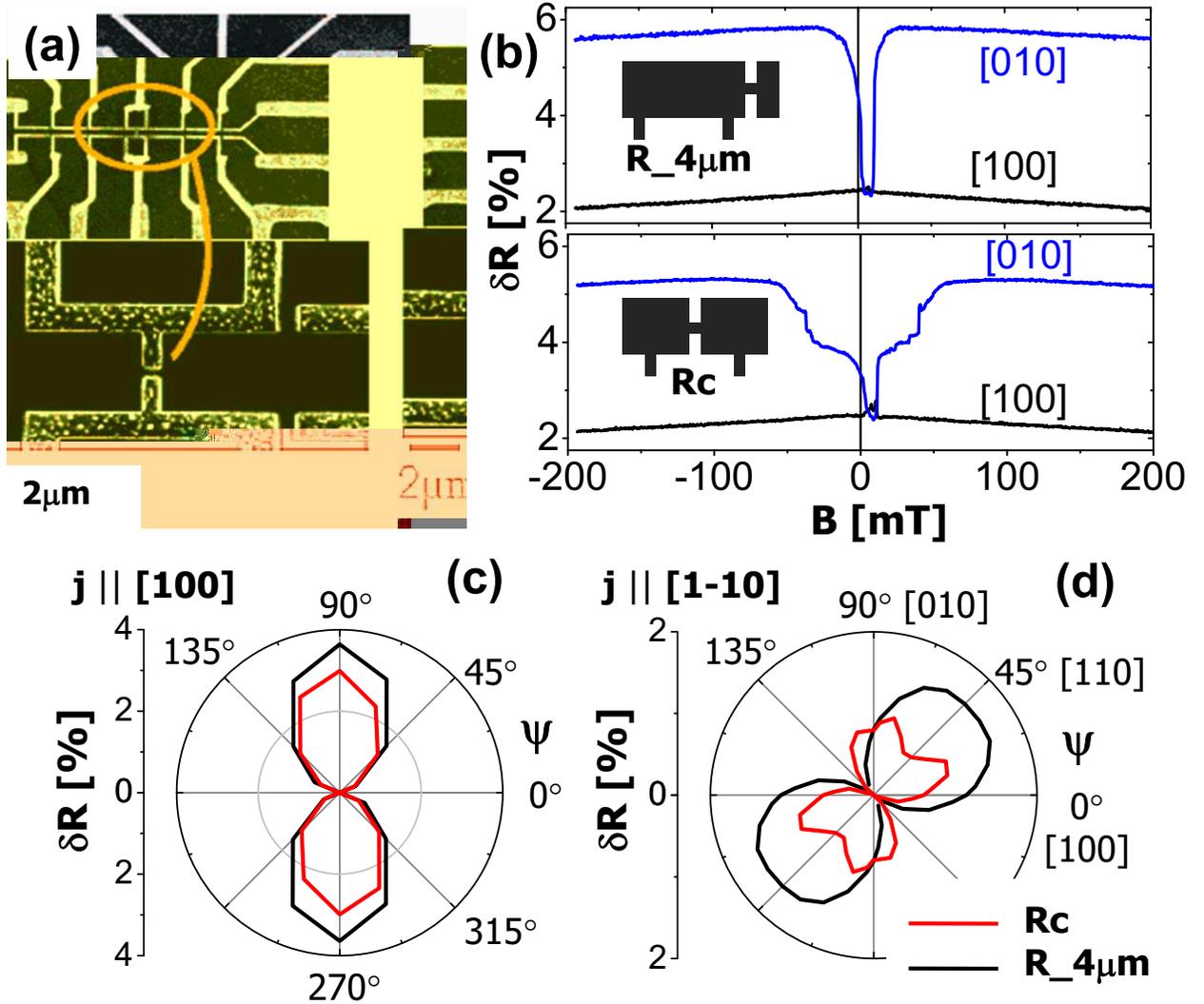}

\caption{(a) Scanning electron micrographs of a 4~$\mu$m wide bar containing a
150~nm wide and 500~nm long constriction patterned from the same wafer
material as in Figs.~\ref{f1} and \ref{f2}.
(b) Resistance variations during in-plane magnetic field sweeps from negative
to positive saturation fields applied along [100] (black) and [010] (blue) directions, measured in the constriction device patterned along the [100] crystallographic axis.
(c,d) AMR measurements in the wider contact (black) and across the constriction (red) in a rotating
saturation field of 4~T for devices patterned along the [100] direction (c) and along
the [110] direction (d). Percentage change
in resitances rather than resistivities are plotted for this non-uniform geometry device; the distinction is not relevant for the discussion of the relative changes in the longitudinal magnetoresistance.}
\label{f3}
\end{figure}

Measurements in the Hall bars discussed above demonstrate that (sub)micrometer lithography of (Ga,Mn)As materials grown under lattice matching strains inevitably produces strain relaxation which may be large enough to significantly modify magnetotransport characteristics of the structure. Lateral micro and nanoconstrictions, utilized in magnetotransport studies of non-uniformly magnetized systems or as pinning centers for domain wall dynamics studies, are an important class of devices for which these effects are highly relevant. In Fig.~\ref{f3} we show data measured in devices consisting of two 4~$\mu$m wide bars patterned from the same (Ga,Mn)As wafer as above along the [1$\overline{1}$0] (or [100]) crystallographic direction and connected by a 150~nm wide and 500~nm long constriction. Magnetic field sweep experiments at a fixed field angle, plotted in panel (b), illustrate a marked increase in the constriction of the anisotropy field along the [100] bar direction  at which magnetization rotates from saturation field orientations towards the easy [100]-axis. Because of the dilute moment nature of the (Ga,Mn)As ferromagnet, shape anisotropy plays only a minor role here and the effect is ascribed to strain relaxation and corresponding changes in the magnetocrystalline anisotropy in the constriction.

AMR measurements in rotating $B=4$~T field shown in Figs.~\ref{f3}(c) and (d) provide further indication of the presence of strong strain relaxation induced magnetocrystalline effects in devices with narrow constrictions.
The comparison between AMRs of the wider contacts and of the constriction shows very similar phenomenology to that of the macroscopic and strain-relaxed microscopic Hall bars discussed in the first part of this section (compare Fig.~\ref{f2} and Figs.~\ref{f3}(c) and (d). We again identify the uniaxial crystalline AMR term in the constriction due to microfabrication which is of the same sign and similar magnitude as observed in the micro Hall bars. Consistency is also found when comparing the character of the AMR curves for the micro Hall-bars and for the constriction devices patterned along different crystallographic directions (see Fig.~\ref{f2} and Figs.~\ref{f3}(c) and (d)).

\section{4. Experiments in (Ga,Mn)As/piezo-transducer hybrid structures}
Lithographic patterning of micro and nanostructures in (Ga,Mn)As provides powerful means for engineering the crystalline AMR components. In this section we show that  further, dynamical control of these effects is achieved in hybrid piezoelectric/(Ga,Mn)As structures.
A 25~nm thick Ga$_{0.94}$Mn$_{0.06}$As epilayer utilized in the study was grown by low-temperature molecular-beam-epitaxy on GaAs substrate and buffer layers \cite{Rushforth:2008_a}.
A macroscopic Hall bar, fabricated in the (Ga,Mn)As wafer by optical lithography, and orientated along the [1$\bar{1}$0] direction, was bonded to the PZT piezo-transducer using a two-component epoxy after thinning the substrate to $150 \pm 10$~$\mu$m by chemical etching. The stressor was slightly misaligned so that a positive/negative voltage produces a uniaxial tensile/compressive strain at $\approx -10^{\circ}$ to the [1$\bar{1}$0] direction.
The induced strain was measured by strain gauges, aligned along the [1$\bar{1}$0] and [110] directions, mounted on a second piece of $150 \pm 10$~$\mu$m thick wafer bonded to the piezo-stressor. Differential thermal contraction of GaAs and PZT on cooling to 50~K produces a measured biaxial, in- plane tensile strain at zero bias of ~$10^{-3}$ and a uniaxial strain estimated to be of the order of $\sim 10^{-4}$ \cite{Habib:2007_a}. At 50~K, the magnitude of the additional strain for a piezo voltage of $\pm 150$~V is approximately $2\times 10^{-4}$.
\begin{figure}[h]
\hspace*{0cm}\includegraphics[width=.8\columnwidth,angle=90]{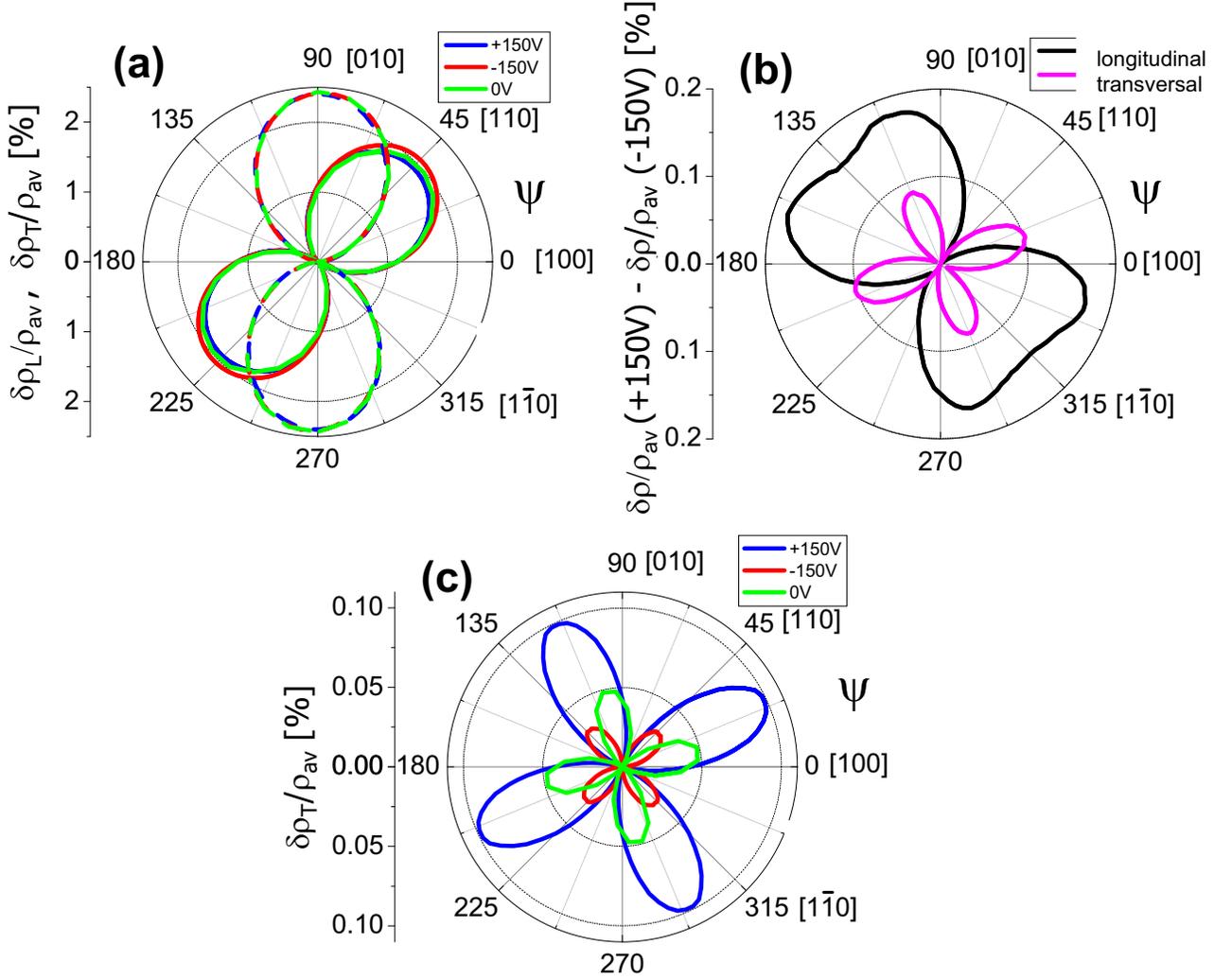}

\caption{(a) The longitudinal (solid curves) and the transverse (dashed curves)
  AMRs for piezo voltages $\pm$150V.  (b) The differences between
  longitudinal and transverse AMRs for piezo voltages  $\pm$150V.  (c) Fourth
  order components of the transverse AMR at piezo voltages $\pm$150V and 0V
  (2nd order components were subtracted as described in the text).  In all
  cases T=50~K and the field of 1~T was rotated in the plane of the (Ga,Mn)As
  layer. As in Fig.s~1-3 we plot better clarity $\delta\rho_L/\rho_{av}\equiv (\rho_L-\rho_{L,min})/\rho_{av}$ and $\delta\rho_T/\rho_{av}\equiv (\rho_T-\rho_{T,min})/\rho_{av}$. Here $\rho_{L(T),min}$ is the minimum (with respect to $\psi$) longitudinal(transverse) resistivity.}
\label{f4}
\end{figure}

Previous measurements \cite{Rushforth:2008_a} of the device identified large changes in the magnetic easy axis orientation induced by the piezoelectric stressor. Here we focus on the effects of the stressor on the magnetotransport coefficients. The AMR measured at 50~K for $\pm 150$~V on the transducer is shown in Fig.~\ref{f4}(a). The modification of the AMR induced by the strain can be extracted by subtracting curves at $\pm 150$~V (see Fig.~\ref{f4}(b)). It is expected that only the crystalline terms are modified; indeed the modification  in the longitudinal resistivity $\rho_L$ is due to the second and fourth order crystalline AMR terms. This is consistent with our previous analysis on unstressed Hall bars where we found that there were second and fourth order crystalline terms representing approximately 10\% of the total AMR. There is also a modification of $\rho_T$ of similar magnitude. This is predominantly due to the fourth order term.

To extract the absolute value of the fourth order term in $\rho_T$ at each voltage we have performed the following analysis: Starting with the raw $\rho_T$ data we subtract any offset due to mixing of $\rho_L$ into the $\rho_T$ signal which may occur as a consequence of small inaccuracies in the Hall bar geometry or small
inhomogeneity in the wafer. This is a correction of approx 0.4\% of the $\rho_L$ signal which should have no significant effect on the subsequent analysis of fourth order terms. (The fourth order components in $\rho_L$ are typically 0.1\%, so the effect on $\rho_T$ would be 0.4\%$\times$0.1\% = 0.0004\%, {\em i.e.}, negligibly small.) We then remove any unintentional antisymmetric (Hall) component from $\rho_T$ by shifting the data by 180$^{\circ}$ and averaging. The second order terms are subsequently removed from $\rho_T$ by shifting the data by 90$^{\circ}$ and averaging. The result of this procedure is plotted in Fig.~\ref{f4}(c).

At 0~V the fourth order component is approximately 0.03\% (peak to trough). At
+150V it is further enhanced to approximately 0.1\% while at -150V the
magnitude is reduced to approximately 0.01\% which is a value similar to the
fourth order term observed after carefully reexamining a (Ga,Mn)As wafer
without the piezo-stressor attached to it \cite{Rushforth:2007_a}. For the
present device, measurements of the magnetic anisotropy indicate that the
application of -150V to the piezo transducer counteracts the uniaxial strain
induced by differential thermal contraction on cooling to return the device
close to the unstrained state \cite{Rushforth:2008_a}. The presence of a
fourth order term in the transverse AMR is allowed under a uniaxial
distortion, see Eq.~(\ref{eq-10b}), but is not expected if only cubic symmetry
is present. The data presented in
Fig.~\ref{f4}(c) clearly demonstrates that the uniaxial strain produced by the
piezo transducer induces a significant fourth order term in the transverse
AMR, which is usually considered to be of insignificant magnitude in the
unstrained wafer. The analysis demonstrates that by applying voltage on the
piezoelectric transducer one can significantly enhance crystalline AMR
components, as compared to the bare (Ga,Mn)As wafer, as well as efficiently
compensate additional strain effects induced by, e.g., different thermal
expansion coefficients in hybrid multilayer structures.

\section{5. Conclusions}

We have demonstrated that beside the previously observed effects on magnetic anisotropies, post-growth strain engineering can be also used to manipulate efficiently the AMR of (Ga,Mn)As. Since magnetic anisotropy is a property of the total energy of the system while AMR reflects quasiparticle scattering rate characteristics \cite{Rushforth:2007_a} there is no straightforward link between the two observations. Experiments and phenomenological analysis of the data have been presented  for two distinct approaches to post-growth strain control: We used the transverse in-plane relaxation of the GaAs/(Ga,Mn)As lattice mismatch strain in
lithographically patterned narrow Hall bars, and a dynamically
controlled strain was induced using a piezo-transducer. Our main results include the observation of AMR
changes due to strain which can be comparable in magnitude to the strongest,
non-crystalline AMR component in bare (Ga,Mn)As, and we have also reported previously undetected
high-order crystalline AMR terms. Finally we have demonstrated that
strain-induced effects can play an important role in magnetoresistance characteristics of (Ga,Mn)As nanoconstrictions.

\section*{Acknowledgements}
We acknowledge support from EU Grant IST-015728, from UK Grant GR/S81407/01,
from Grant Agency and Academy of Sciences and Ministry of Education of the
Czech Republic Grants FON/06/E002, AV0Z10100521, KJB100100802, KAN400100652,
and LC510.

\section*{Appendix -- derivation of phenomenological AMR expressions}

To derive the appropriate AMR expansions for cubic and uniaxially distorted crystals we consider the resistivity tensor in Eq.~(\ref{eq-01}) with the two Hall bars and magnetization vector fixed in space and perform the relevant symmetry operations to the underlying crystal.  (Note that the values of $\psi$ and $\theta$ may
change under
the effect of the symmetry operations since the angles are defined with respect to the crystallographic directions.)
The relevant operations for the cubic crystal are summarized in Tab.~\ref{tab-01}; the last
operation, the invariance under $\psi \to 90^\circ-\psi$ assuming the Hall bars and the crystal fixed, is derived from the microscopic theoretical expression for the AMR \cite{Jungwirth:2002_c}.
The general form of Eq.~(\ref{eq-01}) constrained by these cubic symmetry considerations reads:
%
\begin{equation}
  \hat{\rho} = \hat{\rho}_{cub} = \left(\begin{array}{cc}
   u(\cos^2\psi) &
         \cos\psi\sin\psi \frac{v(\cos^2\psi)+v(\sin^2\psi)}{2}\\
   \cos\psi\sin\psi \frac{v(\sin^2\psi)+v(\cos^2\psi)}{2}
         & u(\sin^2\psi)
               \end{array}\right)  \,.
\label{eq-06}
\end{equation}
%

\begin{table}
\begin{center}
  \begin{tabular}{ll}
    symmetry operation      & implied conditions on $\hat{\rho}$ \\ \hline
    symmetry along $[010]$ &
      $\rho_{11}(\cos\psi,\sin\psi) = \rho_{11}(-\cos\psi,\sin\psi) $\\
    symmetry along $[110]$ &
      $\rho_{11}(\cos\psi,\sin\psi) = \rho_{22}(\sin\psi,\cos\psi) $\\
 &    $\rho_{12}(\cos\psi,\sin\psi) = \rho_{21}(\sin\psi,\cos\psi) $\\
    symmetry along $[1\bar{1}0]$ &
      $\rho_{11}(\cos\psi,\sin\psi) = \rho_{22}(-\sin\psi,-\cos\psi) $\\
    rotation by $90^\circ$ &
      $\rho_{12}(\cos\psi,\sin\psi) =-\rho_{21}(-\sin\psi,\cos\psi) $\\
    invariance under $\psi\to 90^\circ -\psi$ &
      $\rho_{12}(\cos\psi,\sin\psi) = \rho_{12}(\sin\psi,\cos\psi)$ \\
    (fixed crystal)
  \end{tabular}
\end{center}
\caption{Symmetry operations used for a cubic crystal.}
\label{tab-01}
\end{table}

Functions $u$ and $v$ can be expanded in Taylor
series of $\cos^n \psi$ as done in the original work by D\"{o}ring \cite{Doring:1938_a} or, equivalently, in series of $\cos n\psi$. For example for $u$ in Eq.~(\ref{eq-06}) we obtain,
\begin{equation}
u(\cos^2\psi) = a_0 + a_2 \cos 2\psi + a_4 \cos 4\psi + \ldots
\label{eq-07}
\end{equation}
and
\begin{equation}
u(\sin^2\psi)=a_0 - a_2 \cos 2\psi + a_4 \cos 4\psi - \ldots\;.
\label{eq-07b}
\end{equation}
Eqs.~(\ref{eq-06})-(\ref{eq-07b}) together with Eq.~(\ref{eq-02}) yield, after transforming all products of goniometric functions and
recollecting them into sines and cosines of sums of angles, the following structure of the
longitudinal and transverse AMR expressions:
\begin{eqnarray}\label{eq-08a}
  \displaystyle\frac{\Delta\rho_{L}}{\rho_{av}} &=&
          C_C\cos 4\psi + C_{C8}\cos 8\psi + \ldots \\
\nonumber
      && +C_{I}\cos (2\psi-2\theta) + C_{IC} \cos(2\psi+2\theta) + \\
\nonumber
      && +C_{I6}\cos (6\psi-2\theta) + C_{IC6} \cos(6\psi+2\theta) + \\
\nonumber
      && \ldots\,,
\end{eqnarray}
and
\begin{eqnarray}\label{eq-08b}
  \displaystyle \frac{\rho_T}{\rho_{av}} &=& \\
\nonumber
      && +C_{I}\sin (2\psi-2\theta) - C_{IC} \sin(2\psi+2\theta) + \\
\nonumber
      && +C_{I6}\sin (6\psi-2\theta) - C_{IC6} \sin(6\psi+2\theta) + \\
\nonumber
      && \ldots
\end{eqnarray}
Eq.~(\ref{eq-04a},\ref{eq-04b}) in Section~2 are obtained by keeping all terms in
(\ref{eq-08a},\ref{eq-08b}) up to $4\psi$. Note that there is a simple relationship between the longitudinal and transverse AMRs, $\rho_T/\rho_{av}=-\frac{1}{2}(\partial (\Delta\rho_L/\rho_{av})/\partial\theta)$,
which is a consequence of the symmetry $(\hat{\rho})_{ij}=(\hat{\rho})_{ji}$ in Eq.~(\ref{eq-06}).

\begin{table}
\begin{center}
  \begin{tabular}{ll}
    symmetry operation      & implied conditions on $\hat{\rho}$ \\ \hline
    symmetry along $[110]$ &
      $\rho_{11}(\cos\psi,\sin\psi) = \rho_{22}(\sin\psi,\cos\psi) $\\
 &    $\rho_{12}(\cos\psi,\sin\psi) = \rho_{21}(\sin\psi,\cos\psi) $\\
    symmetry along $[1\bar{1}0]$ &
      $\rho_{11}(\cos\psi,\sin\psi) = \rho_{22}(-\sin\psi,-\cos\psi) $\\
 &    $\rho_{12}(\cos\psi,\sin\psi) = \rho_{21}(-\sin\psi,-\cos\psi) $\\
    invariance under $\psi\to 90^\circ -\psi$ &
      $\rho_{12}(\cos\psi,\sin\psi) = \rho_{12}(\sin\psi,\cos\psi)$ \\
    (fixed crystal)
  \end{tabular}
\end{center}
\caption{Symmetry operations used for cubic crystal uniaxially strained along $[110]$.}
\label{tab-02}
\end{table}

Analogous procedure can be applied to cubic crystals with uniaxial strain
along the $[110]$-direction; corresponding symmetry operations are listed in Tab.~\ref{tab-02} and
$\hat{\rho}$ in this case reads,
%
\begin{equation}
  \hat{\rho} = \hat{\rho}_{cub} + \left(\begin{array}{cc}
     t(\cos^2\psi) \cos\psi\sin\psi &
                             \frac{1}{2}[w(\cos^2\psi)+w(\sin^2\psi)]  \\
                        \frac{1}{2}[w(\cos^2\psi)+w(\sin^2\psi)]
&
             t(\sin^2\psi) \cos\psi\sin\psi
               \end{array}\right)\;.
\label{eq-09}
\end{equation}
Eq.~(\ref{eq-09}) yields the following uniaxial AMR terms,
\begin{eqnarray}\label{eq-10a}
   \frac{\Delta \rho_L}{\rho_{av}} &=&
               C_U^{[110]} \sin2\psi +
             C_{U6}^{[110]}\sin 6\psi + C_{U10}^{[110]}\sin 10\psi+\ldots\\
\nonumber
      && + C_{IU}^{[110]} \sin2\theta + \\
\nonumber
&& + C_{IU4+}^{[110]}\sin(4\psi-2\theta) +
     C_{IU4-}^{[110]}\sin(4\psi+2\theta) +\\
\nonumber
      && + C_{IU8+}^{[110]}\sin(8\psi-2\theta)
         + C_{IU8-}^{[110]}\sin(8\psi+2\theta) +\\
\nonumber && \dots
\end{eqnarray}
and
\begin{eqnarray}\label{eq-10b}
    \frac{\rho_T}{\rho_{av}} &=&  \\
\nonumber
      && + C_{IU}^{[110]} \cos2\theta + \\
\nonumber
      && - C_{IU4+}^{[110]}\cos(4\psi-2\theta)
         + C_{IU4-}^{[110]}\cos(4\psi+2\theta) +\\
\nonumber
      && - C_{IU8+}^{[110]}\cos(8\psi-2\theta)
         + C_{IU8-}^{[110]}\cos(8\psi+2\theta) +\\
\nonumber && \dots \,.
\end{eqnarray}
The terms which contain at most $2\psi$ reproduce Eq.~(\ref{eq-11a}).

\begin{table}
\begin{center}
  \begin{tabular}{ll}
    symmetry operation      & implied conditions on $\hat{\rho}$ \\ \hline
    symmetry along $[100]$ &
      $\rho_{11}(\cos\psi,\sin\psi) = \rho_{11}(\cos\psi,-\sin\psi) $\\
&     $\rho_{22}(\cos\psi,\sin\psi) = \rho_{22}(\cos\psi,-\sin\psi) $\\
&     $\rho_{12}(\cos\psi,\sin\psi) =-\rho_{12}(\cos\psi,-\sin\psi) $\\
&     $\rho_{21}(\cos\psi,\sin\psi) =-\rho_{21}(\cos\psi,-\sin\psi) $\\
    symmetry along $[010]$ &
      $\rho_{11}(\cos\psi,\sin\psi) = \rho_{11}(-\cos\psi,\sin\psi) $\\
&     $\rho_{22}(\cos\psi,\sin\psi) = \rho_{22}(-\cos\psi,\sin\psi) $\\
&     $\rho_{12}(\cos\psi,\sin\psi) =-\rho_{12}(-\cos\psi,\sin\psi) $\\
&     $\rho_{21}(\cos\psi,\sin\psi) =-\rho_{21}(-\cos\psi,\sin\psi) $\\
  \end{tabular}
\end{center}
\caption{Symmetry operations used for a cubic crystal uniaxially strained along $[100]$.}
\label{tab-03}
\end{table}

Cubic crystal with uniaxial strain along $[100]$-axis are described by (see Tab.~\ref{tab-03}),
\begin{equation}\hskip-2.5cm
  \hat{\rho} = \left(\begin{array}{cc}
      u(\cos^2\psi)+\Delta u(\cos^2\psi) &
          \sin\psi\cos\psi[v(\cos^2\psi)+\Delta v(\cos^2\psi)] \\
      \sin\psi\cos\psi[v(\sin^2\psi)-\Delta v(\sin^2\psi)] &
          u(\sin^2\psi)-\Delta u(\sin^2\psi)
\end{array}\right)\,.
\label{eq-09b}
\end{equation}
Note that $(\hat{\rho})_{ij}\neq(\hat{\rho})_{ji}$ in this case. Eq.~(\ref{eq-09b}) yields the following uniaxial AMR terms,
%
\begin{eqnarray}\label{eq-12a}
  \frac{\Delta\rho_L}{\rho_{av}} &=&
               C_U^{[100]} \cos2\psi + C_{U6}^{[100]}\cos 6\psi +
                   C_{U10}^{[100]}\cos 10\psi+\ldots\\
\nonumber
      && + C_{IU}^{[100]} \cos2\theta + \\
\nonumber
      && + C_{IU4+}^{[100]}\cos(4\psi-2\theta) +
           C_{IU4-}^{[100]}\cos(4\psi+2\theta) +\\
\nonumber
      && + C_{IU8+}^{[100]}\cos(8\psi-2\theta) +
           C_{IU8-}^{[100]}\cos(8\psi+2\theta) +\\
\nonumber && \dots\,,
\end{eqnarray}
and
\begin{eqnarray}\label{eq-12b}
     \frac{\rho_T}{\rho_{av}}
     &=& +C_{U,T}^{[100]}\sin 2\psi +
      C_{U4,T}^{[100]}\sin 4\psi + C_{U6,T}^{[100]}\sin 6\psi+\ldots \\
\nonumber
      && - C_{IU}^{[100]} \sin2\theta + \\
\nonumber
      && + C_{IU4+}^{[100]}\sin(4\psi-2\theta)
         - C_{IU4-}^{[100]}\sin(4\psi+2\theta) +\\
\nonumber
      && + C_{IU8+}^{[100]}\sin(8\psi-2\theta)
         - C_{IU8-}^{[100]}\sin(8\psi+2\theta) +\\
\nonumber && \dots \,.
\end{eqnarray}
Again the lowest order terms reproduce Eq.~(\ref{eq-11b}).



\end{document}